\documentclass[twocolumn]{aastex631}

\usepackage[utf8]{inputenc}

\shorttitle{Near $K$-Edge Photoionization and Photoabsorption of  Si$^+$, Si$^{2+}$, and Si$^{3+}$}
\shortauthors{Near $K$-Edge Photoionization and Photoabsorption of  Si$^+$, Si$^{2+}$, and Si$^{3+}$}
\begin{document}

\title{Near K-Edge Photoionization and Photoabsorption of Singly, Doubly, and Triply Charged Silicon Ions}

\author{Stefan Schippers}
\affiliation{I. Physikalisches Institut, Justus-Liebig-Universit\"{a}t Gie{\ss}en, Heinrich-Buff-Ring 16, 35392 Giessen, Germany}
\correspondingauthor{Stefan Schippers}
\email{schippers@jlug.de}

\author{Sebastian Stock}
\affiliation{Helmholtz-Institut Jena, Fr\"obelstieg 3, 07743 Jena, Germany}
\affiliation{Theoretisch-Physikalisches Institut, Friedrich-Schiller-Universit\"at Jena, 07743 Jena, Germany}

\author{Ticia Buhr}
\affiliation{I. Physikalisches Institut, Justus-Liebig-Universit\"{a}t Gie{\ss}en, Heinrich-Buff-Ring 16, 35392 Giessen, Germany}

\author{Alexander Perry-Sassmannshausen}
\affiliation{I. Physikalisches Institut, Justus-Liebig-Universit\"{a}t Gie{\ss}en, Heinrich-Buff-Ring 16, 35392 Giessen, Germany}

\author{Simon Reinwardt}
\affiliation{Institut f\"{u}r Experimentalphysik, Universit\"{a}t Hamburg, Luruper Chaussee 149, 22761 Hamburg, Germany}

\author{Michael Martins}
\affiliation{Institut f\"{u}r Experimentalphysik, Universit\"{a}t Hamburg, Luruper Chaussee 149, 22761 Hamburg, Germany}

\author{Alfred M\"{u}ller}
\affiliation{Institut f\"{u}r Atom- und Molek\"{u}lphysik, Justus-Liebig-Universit\"{a}t Gie{\ss}en, Leihgesterner Weg 217, 35392 Giessen, Germany}

\author{Stephan Fritzsche}
\affiliation{Helmholtz-Institut Jena, Fr\"obelstieg 3, 07743 Jena, Germany}
\affiliation{Theoretisch-Physikalisches Institut, Friedrich-Schiller-Universit\"at Jena, 07743 Jena, Germany}

\begin{abstract}
Experimental and theoretical results are presented for  double, triple, and quadruple photoionization of Si$^+$ and Si$^{2+}$ ions and for double photoionization of Si$^{3+}$ ions by a single photon. The experiments employed the photon-ion merged-beams technique at a synchrotron light source. The experimental photon-energy range 1835--1900~eV comprises resonances associated with the excitation of a $1s$ electron to higher subshells and subsequent autoionization. Energies, widths, and strengths of these resonances are extracted from high-resolution photoionization measurements, and the core-hole lifetime of K-shell ionized neutral silicon is inferred. In addition, theoretical cross sections for photoabsorption and multiple photoionization were obtained from large-scale Multi-Configuration Dirac-Hartree-Fock (MCDHF) calculations. The present calculations agree with the experiment much better than previously published theoretical results. The importance of an accurate energy calibration of laboratory data is pointed out. The present benchmark results are particularly useful for discriminating between silicon absorption in the gaseous and in the solid component (dust grains) of the interstellar medium.
\end{abstract}


\section{Introduction}

Silicon is a relatively abundant element in the Universe and, in particular, a major component of interstellar dust.  The binding energy of Si $K$-shell electrons amounts to about 1840~eV \citep{Deslattes2003}. Correspondingly, Si $K$-shell absorption by the interstellar medium (ISM) can be observed by x-ray satellites when directed towards distant x-ray emitting cosmic objects such as x-ray binaries \citep{Rogantini2020}. For the correct determination of the silicon abundance in the ISM it is important to know how much silicon is in the gaseous state and how much is (chemically) bound to dust grains \citep[see, e.g.,][]{Jenkins2009}. Since high-resolution x-ray absorption spectroscopy is sensitive to chemical shifts of characteristic absorption lines, astrophysical models of the ISM have recently been augmented by absorption coefficients of silicon-containing minerals that were obtained from laboratory measurements at a synchrotron light source \citep{Zeegers2017,Zeegers2019}. Here we provide complementary laboratory data for $K$-shell x-ray absorption by atomic  Si$^+$, Si$^{2+}$, and Si$^{3+}$ ions that address the gaseous component of the ISM.

Previous experimental work on the photoabsorption of silicon ions considered only absorption by outer electronic shells \citep{Mosnier2003a,Bizau2009,Kennedy2014}. With the very recent exception of negatively charged Si$^-$ \citep{Perry-Sassmannshausen2021},  $K$-shell x-ray absorption data for silicon ions were hitherto  exclusively obtained from  theoretical calculations \citep{Verner1993a,Palmeri2008,Witthoeft2011,Kucas2012,Kucas2015,Hasoglu2021}. Because of the many-body nature of the theoretical problem the theoretical calculations have to resort to approximations yielding results that bear (usually unknown) uncertainties. In this situation, benchmarking by laboratory experiments \citep{Schippers2020c} is vital for arriving at sufficiently accurate results that allow one, in the present context,  to reliably discriminate between the gaseous and solid components of the ISM.

Similar to our previous work on  $L$-shell ionization of Fe$^+$, Fe$^{2+}$, Fe$^{3+}$ and Ar$^+$ ions \citep{Schippers2017,Beerwerth2019,Schippers2021,Mueller2021b}, we have investigated ionization of silicon ions in different charge states. Here, we present experimental and theoretical data for $m$-fold photoionization of Si$^{q+}$ ions with primary charge state $q$ leading to the production of Si$^{r+}$ ions with product charge-state $r=q+m$
\begin{equation}\label{eq:multi}
h\nu + \mathrm{Si}^{q+} \to \mathrm{Si}^{r+}+m\mathrm{e}^-.
\end{equation}
The paper is organized as follows. Section \ref{sec:exp} discusses experimental issues, in particular the role of metastable ions in the silicon ion beams and the energy calibration of the photon energy scale. Section \ref{sec:theo} describes  calculations  of the absorption cross-sections and of the complex deexcitation cascades that set in after the initial creation of a $K$-shell hole. The results are presented and discussed in section \ref{sec:res} emphasizing their  relevance for the identification of gaseous silicon in the ISM. The concluding section \ref{sec:conc} briefly summarizes the findings and provides an outlook on future directions of laboratory astrophysics in the area of atomic inner-shell absorption.

\section{Experiment}\label{sec:exp}

The measurements were carried out at the synchrotron light source PETRA\,III operated by DESY in Hamburg, Germany. More specifically, the photon-ion merged-beam technique \citep[recently reviewed by][]{Schippers2016} was employed using the PIPE end-station \citep{Schippers2014,Mueller2017,Schippers2020} at the photon-beamline P04 \citep{Viefhaus2013}.  The beamline's monochromator  is equipped with two diffraction gratings with different rulings of  400 and 1200 lines/mm. The latter was used for the present experiment because it offers the highest photon flux in the photon-energy region of interest. The experimental procedures have been described previously \citep{Schippers2014,Mueller2017}. Therefore, we only provide the details here that are relevant for the presently-studied ion species.

Beams of Si$^{q+}$ ions ($q$=1--3) were produced by evaporating SiO powder from an electrically heated oven into an electron-cyclotron-resonance (ECR) ion source.  The ion source was operated on a positive potential of 6~kV such that positively charged ions were accelerated from the ion source towards the electrically grounded ion beam-line. Downbeam, a dipole electromagnet served for selecting the desired ion species according to their mass-to-charge ratio. Subsequently, the ion beam was collimated and centered onto the counter-propagating photon beam by tuning electrostatic deflectors and lenses appropriately.  Typical ion currents in the interaction region were  4.5~nA of $^{28}$Si$^+$ and 2.5~nA of $^{28}$Si$^{3+}$. The $^{28}$Si$^{2+}$ ion current was significantly higher (20~nA) since it was heavily contaminated with, e.g.,  N$^+$ or CO$^{2+}$ ions, which have the same mass-to-charge ratio as $^{28}$Si$^{2+}$. Similarly, it cannot be excluded that the Si$^+$ beam was contaminated by N$_2^+$ or CO$^+$ ions. It has to be pointed out that the product ions with charge states increased by 1, 2, 3, $\ldots$ units can all be individually separated by the dipole magnet (the demerger) in front of the ion detector so that each final channel could be individually investigated.

Relative cross sections for multiple photoionization (cf.~Equation~\ref{eq:multi}) were determined over a preselected range of photon energies by normalizing the photon-energy dependent product-ion count rate on the photon flux and on the ion-current, which were measured with a calibrated photodiode and a Faraday cup, respectively. For the Si$^{3+}$ primary ions, the relative cross sections were put on an absolute scale by additionally accounting for the separately measured geometrical beam overlap \citep[for details see][]{Schippers2014,Mueller2017}. The uncertainty of the experimental absolute cross-section scale amounts to $\pm$15\% \citep{Schippers2014}.

The measurement of absolute cross sections requires a careful adjustment of the mutual overlap of the photon and ion beams. This procedure was only applied to the Si$^{3+}$ ion beam. An experimental determination of absolute cross sections for the multiple ionization of Si$^+$ and Si$^{2+}$ primary ions was not possible because of the unknown fractions of contaminating ions in these ion beams. Therefore, the Si$^{+}$ and  Si$^{2+}$ cross sections were normalized to theoretical absorption cross sections as explained below.

\subsection{Metastable primary ions }\label{sec:meta}

The Si$^{q+}$ ion beams contained ground-level ions and possibly also unknown fractions of long-lived metastable ions with lifetimes longer than the ions' flight time (a few microseconds) through the apparatus.   For aluminium-like Si$^+$, where we assume a statistical population of the experimentally unresolved two fine-structure components of the $3s^2\,3p\;^2P$ ground term, the long-lived metastable  levels are the $3s\,3p^2\;^4P_J$ levels with excitation energies of $\sim$5.3~eV  and lifetimes in the range  0.13--1.2~ms \citep{FroeseFischer2006}.   Magnesium-like Si$^{2+}$ has long-lived $3s\,3p\;^3P_J$ excited levels with excitation energies of $\sim$6.6~eV and lifetimes of 58~$\mu$s ($J=1$), 7.8~s ($J=2$), and  $\infty$ ($J=0$)  with regard to single-photon emission \citep{FroeseFischer2006}.  Sodium-like Si$^{3+}$ does not have long-lived singly excited metastable levels. However, the core-excited $2p^5\,3s\,3p\;^4D_{7/2}$ level  is metastable against autoionization with a lifetime in the 1--10 $\mu$s range \citep{Howald1986a}. Its excitation energy is 106~eV \citep{Schmidt2007b}.

The fractional populations of metastable excited levels depend on their excitation energies and on the plasma conditions in the ion source. In outer-shell photoionization experiments, their presence is often revealed by characteristic threshold and resonance features in the measured photoionization cross-sections. By comparing experimental and theoretical photoionization cross-sections \citet{Kennedy2014} inferred a   $3s\,3p^2\;^4P$ fraction of about 10\% for a Si$^+$ ion beam that was also produced with an ECR ion source. Similarly, \citet{Mosnier2003a} arrived at a 2--3\% $3s\,3p\;^3P$ fraction of their Si$^{2+}$ ion beam. In the present inner-shell photoionization measurements we did not recognise any specific cross-section features that can be attributed to the presence of metastable levels. Nevertheless, such metastable levels cannot be excluded for the present Si$^+$ and Si$^{2+}$ beams similar as in the previous experiments by \citet{Kennedy2014} and \citet{Mosnier2003a}.

In view of its high excitation energy, one does not expect a significant population of the autoionizing Si$^{3+}$($2p^5\,3s\,3p\;^4D_{7/2}$) metastable level. However, even a tiny contamination of the Si$^{3+}$ beam can lead to a noticeable production of Si$^{4+}$ ions via autoionization. In the present experiment, this created a large background in the Si$^{3+}$ single-ionization channel, such that no meaningful experimental result could be obtained for the production of Si$^{4+}$ by photoionization of Si$^{3+}$. We conclude that, apart from this  limitation, metastable primary ions do not play a significant role in the present study.

\subsection{Photon-energy calibration}\label{sec:cal}

\begin{figure}[t]
\includegraphics[width=\linewidth]{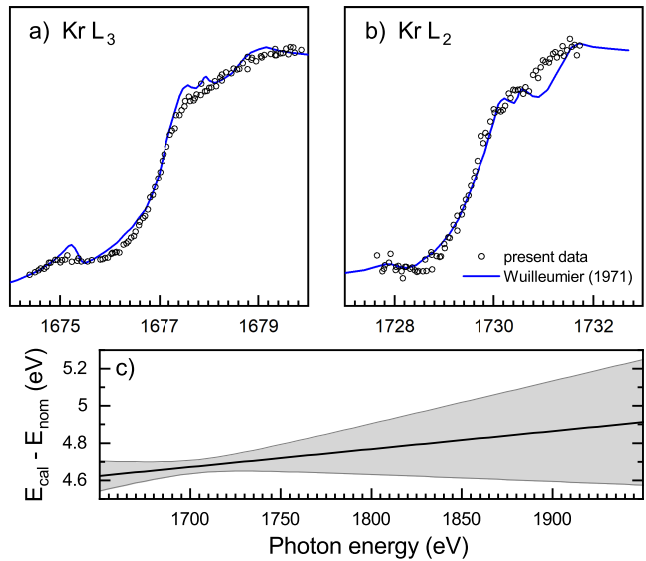}
\caption{\label{fig:cal}Calibration of the photon-energy scale by shifting the present krypton absorption data (symbols) to the experimental results of \citet{Wuilleumier1971} (full lines, digitized) for the $L_3$  and $L_2$  absorption edges displayed in panels a) and b), respectively. The full line in panel c) is the extrapolated difference between calibrated photon energies E$_\mathrm{cal}$ and nominal photon energies $E_\mathrm{nom}$ for the entire energy-range of interest. The gray shaded area visualizes the associated uncertainties as determined from error propagation (see text). In all panels, the abscissae are the calibrated photon energy.}
\end{figure}

The photon-energy scale was calibrated by a separate measurement of the  krypton $L_3$ and $L_2$ absorption edges at about 1677~eV  and 1730~eV (Figure~\ref{fig:cal}) using a combination of a gas jet and a photoelectron spectrometer \citep[see][for details]{Mueller2017,Mueller2018c}. The re\-com\-mended values for the Kr $L_3$ and $L_2$  threshold energies are 1679.07(39) and 1730.90(50)~eV \citep{Deslattes2003}. Unfortunately, it is not clear where to read these values off the measured absorption curves.  Therefore, we have applied energy shifts to our measured data  such that the presently measured krypton absorption edges line up with the corresponding absorption curves of \citet{Wuilleumier1971} as displayed in Figure~\ref{fig:cal}.

\citet{Wuilleumier1971} provided wavelengths in Cu~x~units referring to the Cu $K\alpha_1$ line. For the conversion of these units to electronvolts, the CODATA 2018 set of fundamental physical constants was used \citep{Tiesinga2021}. The resulting conversion factor agrees with the one provided earlier by \citet{Deslattes2003} within its negligible (in the present context) uncertainty.

Absorption at the Kr $L_3$ and $L_2$  edges was also measured by others. The results of \citet{Kato2007} differ from those  of \citet{Wuilleumier1971} by $\sim$1.25 and $\sim$0.55~eV, respectively. \citet{Kato2007} calibrated their energy scale to the Kr $2p_{3/2}^{-1}5p$ excitation energy in the vicinity of the $L_3$ threshold as determined by \citet{Ibuki2002}. These latter authors  did not provide an estimate for the uncertainty of their energy scale. The same holds for the Kr $L_3$  absorption measurements of \citet{Schmelz1995}. \citet{Nagaoka2000} and \citet{Okada2005} calibrated their energy scales by referring to the work of \citet{Wuilleumier1971} as is also done here.

The energy shifts applied to our absorption curves were  4.65 and 4.7~eV for the $L_3$ edge and the $L_2$ edge, respectively.  We attribute a read-off uncertainty of 0.05~eV to this calibration method.  For the extrapolation to higher energies we used a linear fit through the two calibration points.  By this extrapolation, the read-off error propagates to energy uncertainties of  $\pm$0.2~eV at 1840~eV  and of $\pm$0.3~eV  at 1920~eV (Figure~\ref{fig:cal}c). Taking the 0.5~eV uncertainty of the recommended $L_2$-threshold energy \citep{Deslattes2003}  and a 0.2 eV uncertainty associated with imperfections of the photon beamline  (see below) additionally into account, we arrive at an uncertainty of $\pm$1~eV of the present calibrated photon-energy scale  for the  photon energies in the current experimental range of 1835--1900~eV.

In calibrating the photon-energy scale of the silicon ions, we have additionally considered the Doppler shift that is caused by the unidirectional movement of the ions in the Si$^{q+}$ ion beams  \citep[see, e.g.,][for details]{Schippers2014}. This does not introduce any significant additional uncertainty of the photon-energy calibration.

\section{Theory}\label{sec:theo}

\begin{deluxetable}{lll}
\tablecaption{\label{tab:Si1theo}Configurations that are taken into account  in the present calculations of the deexcitation cascades following the direct $1s$ ionization and the $1s\to3p$ excitation of Si$^+$. All in all,  nearly a quarter million of level-to-level transition rates had to be calculated to describe the cascade processes.}
\tablehead{
    \colhead{Ion} &
    \colhead{$1s$ ionization}&
    \colhead{$1s\to3p$ excitation}
    }
\startdata
Si$^{+}$   &   &  $1s^{1}\,2s^2\,2p^6\,3s^2\,3p^2$   \\
 & & \\
Si$^{2+}$  &  $1s^{2}\,2s^2\,2p^6\,3s^2$           &  $1s^{2}\,2s^2\,2p^6\,3s^2$   \\
                 &  $1s^{2}\,2s^2\,2p^6\,3s^1\,3p^1$ &  $1s^{2}\,2s^2\,2p^6\,3s^1\,3p^1$   \\
                 &  $1s^{2}\,2s^2\,2p^5\,3s^2\,3p^1$ &  $1s^{2}\,2s^2\,2p^6\,3p^2$   \\
                 &  $1s^{2}\,2s^2\,2p^5\,3s^1\,3p^2$ &  $1s^{2}\,2s^2\,2p^5\,3s^2\,3p^1$   \\
                 &  $1s^{1}\,2s^2\,2p^6\,3s^2\,3p^1$ &  $1s^{2}\,2s^2\,2p^5\,3s^1\,3p^2$   \\
                 &  &  $1s^{2}\,2s^1\,2p^6\,3s^2\,3p^1$   \\
                 &  &  $1s^{2}\,2s^1\,2p^6\,3s^1\,3p^2$   \\
                 &  &  $1s^{2}\,2s^2\,2p^4\,3s^2\,3p^2$   \\
                 &  &  $1s^{2}\,2s^1\,2p^5\,3s^2\,3p^2$   \\
                 &  &  $1s^{2}\,2p^6\,3s^2\,3p^2$   \\
 & & \\
Si$^{3+}$  & $1s^{2}\,2s^2\,2p^6\,3s^1$              &  $1s^{2}\,2s^2\,2p^6\,3s^1$   \\
                 &  $1s^{2}\,2s^2\,2p^6\,3p^1$             &  $1s^{2}\,2s^2\,2p^6\,3p^1$   \\
                 & $1s^{2}\,2s^2\,2p^5\,3s^2$              &  $1s^{2}\,2s^2\,2p^5\,3s^2$   \\
                 &  $1s^{2}\,2s^2\,2p^5\,3s^1\,3p^1$  &  $1s^{2}\,2s^2\,2p^5\,3s^1\,3p^1$   \\
                 & $1s^{2}\,2s^1\,2p^6\,3s^2$              &  $1s^{2}\,2s^2\,2p^5\,3p^2$   \\
                 & $1s^{2}\,2s^1\,2p^6\,3s^1\,3p^1$   &  $1s^{2}\,2s^1\,2p^6\,3s^2$   \\
                 & $1s^{2}\,2s^2\,2p^4\,3s^2\,3p^1$   &  $1s^{2}\,2s^1\,2p^6\,3s^1\,3p^1$   \\
                 & $1s^{2}\,2s^1\,2p^5\,3s^2\,3p^1$   &  $1s^{2}\,2s^1\,2p^6\,3p^2$   \\
                 & $1s^{2}\,2p^6\,3s^2\,3p^1$              &  $1s^{2}\,2s^2\,2p^4\,3s^2\,3p^1$   \\
                 &  &  $1s^{2}\,2s^2\,2p^4\,3s^1\,3p^2$   \\
                 &  &  $1s^{2}\,2s^1\,2p^5\,3s^2\,3p^1$   \\
                 &  &  $1s^{2}\,2s^1\,2p^5\,3s^1\,3p^2$   \\
 & & \\
Si$^{4+}$  &  $1s^{2}\,2s^2\,2p^6$                      &  $1s^{2}\,2s^2\,2p^6$   \\
                 & $1s^{2}\,2s^2\,2p^5\,3s^1$             &  $1s^{2}\,2s^2\,2p^5\,3s^1$   \\
                 & $1s^{2}\,2s^2\,2p^5\,3p^1$            &  $1s^{2}\,2s^2\,2p^5\,3p^1$   \\
                 &  $1s^{2}\,2s^1\,2p^6\,3s^1$            &  $1s^{2}\,2s^1\,2p^6\,3s^1$   \\
                 & $1s^{2}\,2s^1\,2p^6\,3p^1$            &  $1s^{2}\,2s^1\,2p^6\,3p^1$   \\
                 &  $1s^{2}\,2s^2\,2p^4\,3s^2$            &  $1s^{2}\,2s^2\,2p^4\,3s^2$   \\
                 &  $1s^{2}\,2s^2\,2p^4\,3s^1\,3p$      & $1s^{2}\,2s^2\,2p^4\,3s^1\,3p^1$    \\
                 & $1s^{2}\,2s^1\,2p^5\,3s^2$             &   $1s^{2}\,2s^2\,2p^4\,3p^2$   \\
                 & $1s^{2}\,2s^1\,2p^5\,3s^1\,3p^1$   &    \\
 & & \\
Si$^{5+}$  & $1s^{2}\,2s^2\,2p^5$  &  $1s^{2}\,2s^2\,2p^5$  \\
                 & $1s^{2}\,2s^1\,2p^6$ & \\
                 & $1s^{2}\,2s^2\,2p^4\,3s^1$ &
\enddata
\end{deluxetable}

The present theoretical calculations were performed in the framework of the Multi-Configuration Dirac-Hartree-Fock (MCDHF) method by applying the  \textsc{Grasp} \citep{Joensson2007a} and \textsc{Ratip} \citep{Fritzsche2012a} atomic structure codes for the calculation of atomic energy levels and transition rates.
In particular, the Auger cascades were analyzed and calculated that follow the creation of a $K$-shell hole in initially Si$^{q+}$ ions by direct $1s$ ionisation or by $1s$ $\to$ $np$ excitation.  These cascade computations \citep{Fritzsche2021} support quantitative estimates for the distribution of Si$^{r+}$ product ions (cf.~Equation~\ref{eq:multi}), similar to earlier studies for other atoms  and ions \citep{Stock2017,Beerwerth2017,Buth2018,Beerwerth2019,Schippers2017,Schippers2021}.

In these cascade computations, the Si$^{q+}$ ions  were assumed to be initially in their respective ground configuration with statistically populated fine-structure levels. Cross sections for the direct photoionization
of these ions were obtained for each atomic subshell, while the resonant $1s \to np$ photoexcitation was considered for all shells with $3\leq n \leq 7$ for Si$^+$ and $3 \leq n \leq 8$ for Si$^{2+}$ and Si$^{3+}$. From the incoherent summation of all these direct-photoionization and resonant-excitation contributions, the total photoabsorption cross-sections were determined for Si$^+$, Si$^{2+}$, and Si$^{3+}$ ions.

\begin{figure*}
\centering\includegraphics[width=0.9\textwidth]{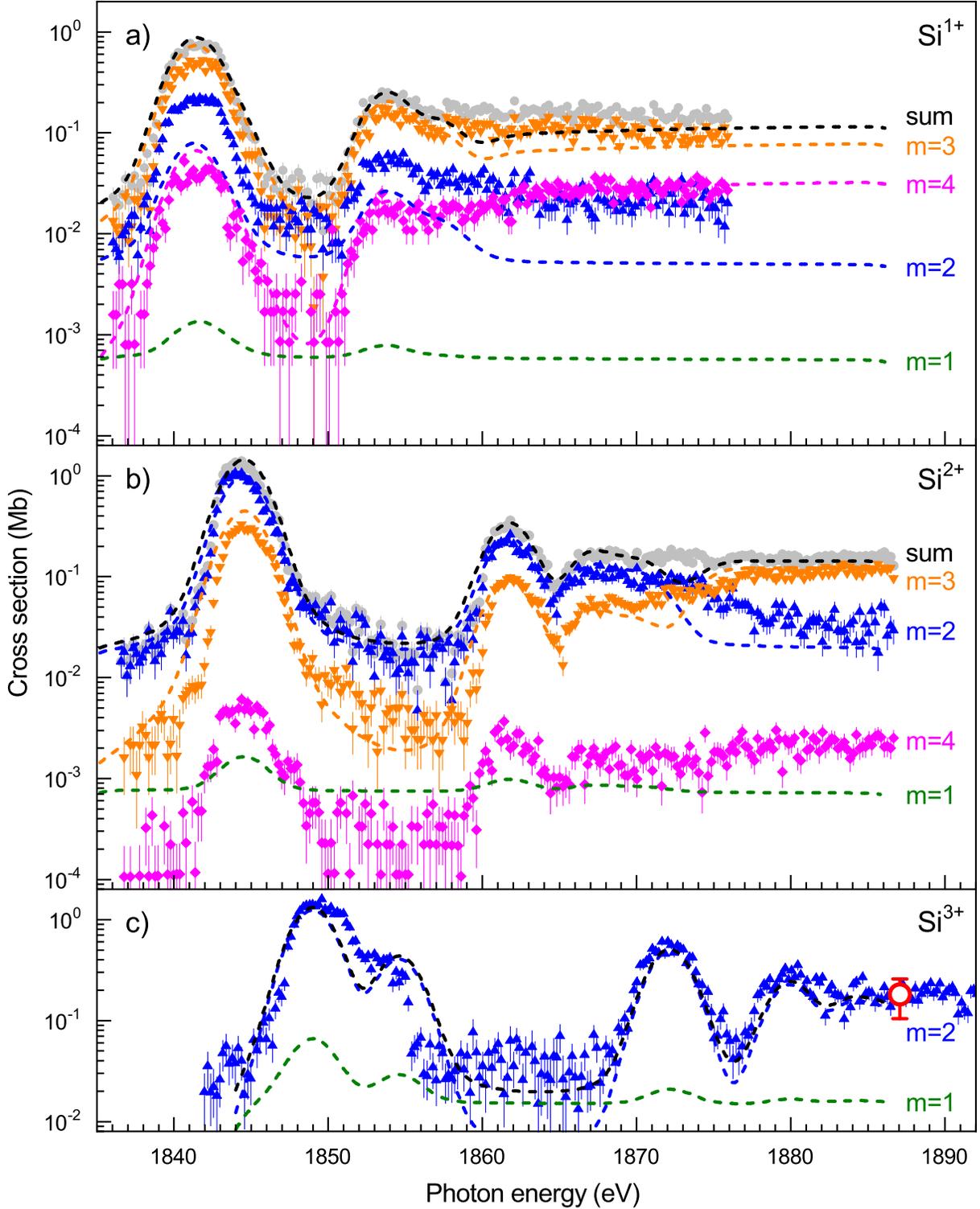}
\caption{\label{fig:all} Experimental cross sections for double (blue triangles up, $m$=2), triple (orange triangles down, $m$=3), and quadruple (magenta diamonds, $m$=4) photoionization of a) Si$^{+}$, b) Si$^{2+}$, and c) Si$^{3+}$. The grey circles in panels a) and b) represent the sum (Equation~\ref{eq:sum}) of the cross sections $\sigma_{1,\Sigma}=\sigma_{1,3}+\sigma_{1,4}+\sigma_{1,5}$ and $\sigma_{2,\Sigma}=\sigma_{2,4}+\sigma_{2,5}+\sigma_{2,6}$, respectively. For Si$^{3+}$, only the double ionization cross-section $\sigma_{3,5}$ was measured. The open circle at $\sim$1887~eV in panel c) represents the  measured absolute data point. Its error bar comprises the statistical uncertainty and a 15\% systematic uncertainty. The dashed curves are the present theoretical results for single ionization (olive), double ionization (blue), triple ionization (orange), and quadruple ionization (magenta). The black dashed curves are the corresponding  cross-section sums. The theoretical cross sections were convolved with a Gaussian with a FWHM of $\Delta E= 3$~eV corresponding to the experimental photon energy spread and shifted by a) $-1.7$~eV, b) $-1.8$~eV, and c) $-1.2$~eV  in order to line up the theoretical resonance features with the experimental ones. Single ionization could not be measured due to the low cross sections and comparatively high background levels, particulary for Si$^{3+}$ where a strong background was caused by autoionizing metastable levels (see Section~\ref{sec:meta}).}
\end{figure*}

While different autoionization processes might contribute to the electron emission from inner-shell excited ions, only those single-step Auger processes were taken into account into the cascade, that are
energetically allowed in a configuration-averaged representation of the atomic fine-structure. This simplifies the cascade computations and makes them tractable, although further contributions from the simultaneous excitation (shake-up) or ionization (shake-off) of  another electron were found essential as well for accurately predicting  the final charge-state distributions of light and especially negatively-charged ions \citep{Mueller2015a,Schippers2016a,Perry-Sassmannshausen2020}.  For multiply-charged ions, however, these shake processes are typically suppressed, though not always negligible. Table~\ref{tab:Si1theo} lists all those electron configurations that were included into the present cascade computations for Si$^+$  ions. The number of cascade steps, i.e., the so-called depth of the cascade, were taken to be 4 for Si$^+$, 3 for Si$^{2+}$ , and 2 for Si$^{3+}$. For the analysis of the cascades, therefore, the highest product-ion charge state was $r=5$ (cf.~Equation~\ref{eq:multi}), while higher charge states were not accessible within the given approach and by excluding autoionization processes with an additional excitation or ionization of electrons.

 Theoretical cross sections for multiple photoionization were eventually obtained by multiplying the calculated absorption cross sections with the branching ratios from the cascade computations. These computations were carried out separately for the different $1s$ core ionized and $1s\to np$ core excited configurations. For the comparison with the experimental results, the theoretical cross sections were convolved with a Gaussian with the full width at half maximum (FWHM) corresponding to the experimental photon energy spread $\Delta E$.

\section{Results and discussion}\label{sec:res}

\subsection{Cross sections for multiple ionization}

Figure \ref{fig:all} provides an overview over the measured and calculated cross sections $\sigma_{q,r}$  for multiple photoionization of Si$^{q+}$ ions (cf.~Equation~\ref{eq:multi}) with primary charge states $q=1,2,3$ together with the cross-section sums
\begin{equation}\label{eq:sum}
\sigma_{q,\Sigma} = \sum_r\sigma_{q,r}.
\end{equation}
The experimental energy ranges comprise the thresholds for direct ionization of one $K$-shell electron.  The experimental photon-energy spread was $\Delta E\approx 3$~eV. This is sufficient for resolving the individual $1s \to np$ resonance groups for $n=3 $ and $n=4$. For Si$^{3+}$, even the $1s\to5p$ resonance group can be discerned.

The logarithmic cross-section scales cover up to  four orders of magnitude.  For Si$^+$, the main ionization channel is triple ionization. For Si$^{2+}$, triple ionization dominates only for energies above $\sim$1875~eV where direct $K$-shell ionization becomes energetically possible.  At lower energies, double ionization is stronger. For Si$^{3+}$, double ionization is the  dominating ionization channel in the entire experimental photon-energy range. The single-ionization channel could not be measured as explained above (Section~\ref{sec:meta}). Other Si$^{3+}$  ionization channels were scrutinized, but no detectable signal was found. This indicates that the associated cross sections are small as is also corroborated by our theoretical calculations.

Our theoretical ab-initio cross sections are in  remarkable agreement with the measured ones considering the simplifications that had to be applied in order to make the  calculations of the complex deexcitation cascades tractable. We like to point out that  the calculations were performed independently of the measurements.

The experimental and theoretical resonances line up if energy shifts of $-1.7$, $-1.8$, and $-1.2$~eV are applied to the Si$^+$, Si$^{2+}$, and Si$^{3+}$ cross sections, respectively. These differences between our theoretical and experimental resonance positions are less than a factor of 2 larger than the 1~eV uncertainty of the experimental energy scale.
Our theoretical calculations do not account for any of the metastable primary ions discussed in Section~\ref{sec:meta}. The comparison between the theoretical and the experimental resonance structures suggests that metastable ions do indeed not play a significant role in the present study as already concluded above.

Only the theoretical and experimental cross-sections  $\sigma_{3,5}$ can be directly compared on an absolute scale. The mutual agreement is within the $\pm$15\% experimental uncertainty over nearly the entire experimental energy range. The largest discrepancy between the experimental and present theoretical cross sections concerns the (small) Si$^+$ double-ionization cross-section which is underestimated by a factor of $\sim$2. At energies between 1845 and 1858~eV, slight discrepancies occur which are due to a limited ability of the theory to correctly describe the experimentally observed resonance structure. There are no such obvious discrepancies for Si$^+$ and Si$^{2+}$.

We conclude that the present theoretical approach is capable of reliably predicting the multiple photoionization cross-sections of low-charged silicon ions. Similar accuracy can be expected for neighboring elements from the periodic table. The benchmarking by our experimental results leads to rather small corrections of the theoretical resonance energies which are only slightly larger than the uncertainty of the present experimental photon-energy scale.

\subsection{Absorption cross-sections}

\begin{figure}[t]
\includegraphics[width=\linewidth]{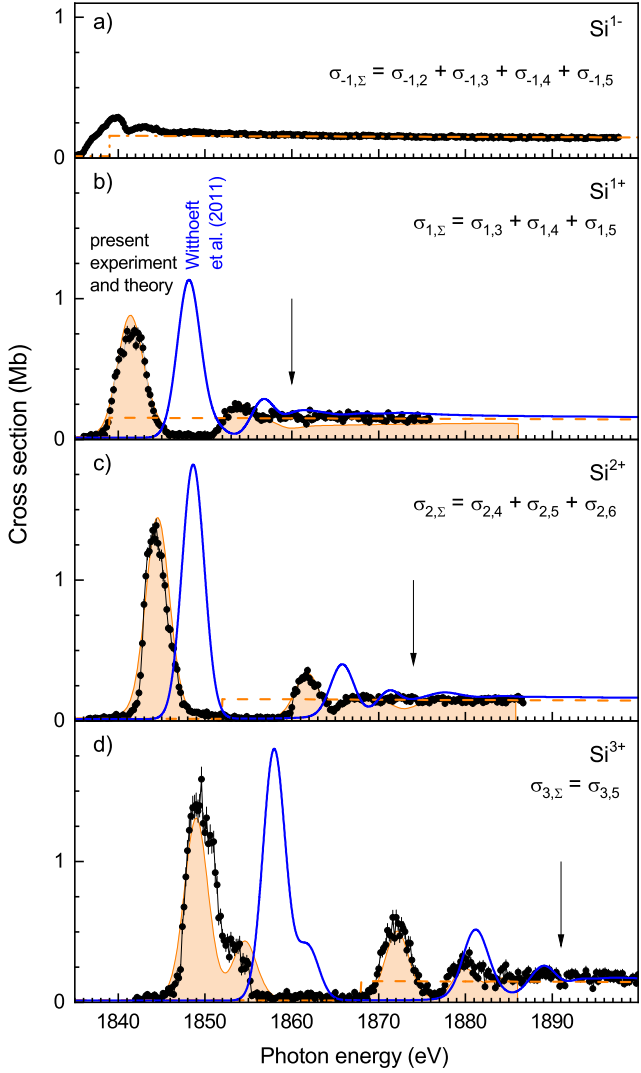}
\caption{\label{fig:abs} Panel a): Experimental cross-section sum (symbols) for negatively charged silicon ions \citep{Perry-Sassmannshausen2021}. The dash-dotted curve is the photoabsorption cross section for neutral silicon atoms as recommended by \citet{Henke1993}. Panels b), c), and d):  Experimental cross-section sums (symbols) from Figure~\ref{fig:all} for  Si$^+$, Si$^{2+}$, and Si$^{3+}$, respectively,  in comparison with corresponding present theoretical photoabsorption cross-sections (shaded curves) and those of \citet[][dashed curves]{Verner1993a} and \citet[][full blue curves]{Witthoeft2011}. For the comparison, the theoretical cross sections were convolved with a Gaussian with a FWHM of 3 eV in order to account for the experimental photon energy spread. In addition, the present theoretical cross sections were shifted by $-1.7$~eV, $-1.8$~eV, and $-1.2$~eV, respectively,  in order to line up the theoretical resonance features with the experimental ones. The vertical arrows mark the present theoretical thresholds for direct $K$-shell ionization with the above energy shifts applied.\\~\\}
\end{figure}

For Si$^+$ and Si$^{2+}$, the experimental cross-section sums $\sigma_{q,\Sigma}$ were put on absolute scales by multiplying all individual cross sections $\sigma_{q,r}$ for a given primary charge state $q$ by the same factor such that the sums line up with the respective theoretical absorption cross-sections of \citet[][Figure~\ref{fig:abs}b--\ref{fig:abs}d]{Verner1993a}. We have applied the same approach already earlier to the photoabsorption of negatively charged Si$^-$ ions \citep[][]{Perry-Sassmannshausen2021} where we used the theoretical absorption cross section for neutral silicon by \citet{Henke1993} as a reference. These data are shown in Figure~\ref{fig:abs}a for comparison.  Apparently, the cross sections  for non-resonant absorption are nearly independent of the primary charge state. At 1890~eV the cross sections  for Si$^0$ \citep{Henke1993} and  for Si$^+$, Si$^{2+}$, and Si$^{3+}$  \citep{Verner1993a} all amount to about 0.15~Mb.

For energies where there are no signatures from resonant processes,  our theoretical cross-section sums for Si$^+$, Si$^{2+}$, and Si$^{3+}$ agree with the  results of \cite{Verner1993a}, which do not account for resonant photoabsorption.  The same is true for the more recent theoretical results of \cite{Witthoeft2011}, who also did include resonant photoabsorption in their calculations. However, their resonance positions deviate significantly by up to $\sim$9~eV from the measured ones. Our present energy-shifted (by less than 2~eV, see above) theoretical results fit much better to the experimentally observed resonance positions than the results of  \cite{Witthoeft2011}.

Another significant discrepancy between the present and the previous results concerns the thresholds for direct $K$-shell photoionization. Our theoretical results are 1860, 1874, and 1891~eV for Si$^+$, Si$^{2+}$, and Si$^{3+}$, respectively, with the above mentioned energy shifts applied.  As can be seen from Figure~\ref{fig:abs}, the threshold energies as predicted by \citet{Verner1993a} are lower by more than 20~eV.  This shows  that our experimental benchmarks are vital for arriving at an accurate positioning of the various absorption features. From the comparison with the theoretical absorption cross sections we conclude that, within the present  experimental uncertainties, the  experimental cross-section sums $\sigma_{q,\Sigma}$ represent the Si$^{q+}$ photoabsorption cross-sections.

\subsection{High-resolution measurements of resonances}

\begin{figure}[b]
\centering{\includegraphics[width=0.95\linewidth]{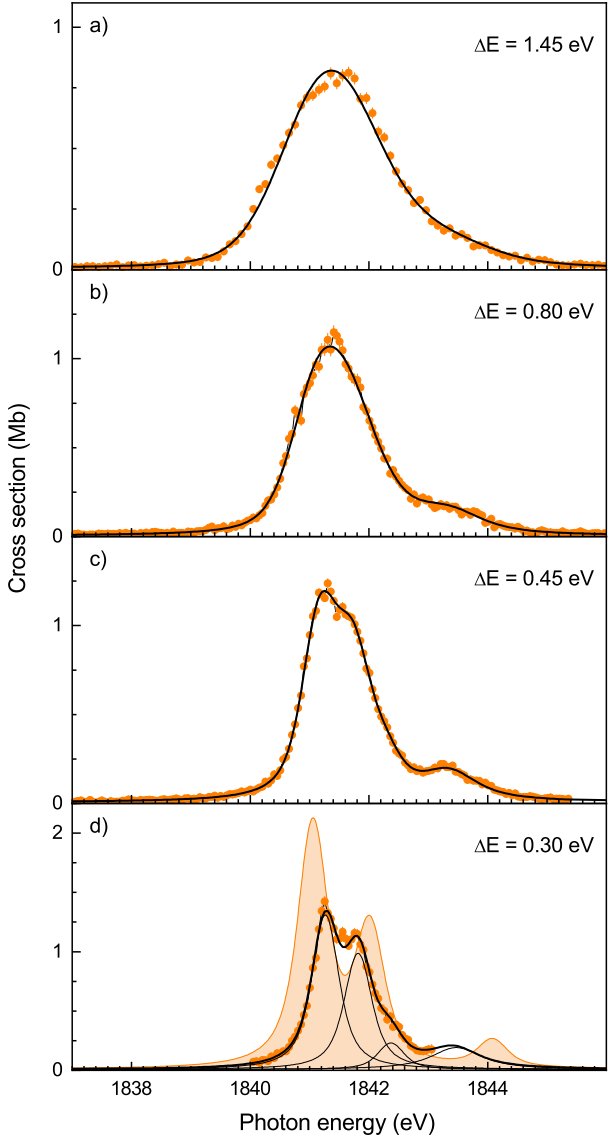}}
\caption{\label{fig:Si1fit} Measured cross sections $\sigma_{1,4}$ for triple photoionization of Si$^{+}$ ions using different photon-energy spreads $\Delta E$ of a) 1.45~eV (500~$\mu$m), b) 0.80~eV (250~$\mu$m), c) 0.45~eV (100~$\mu$m) and d) 0.30~eV (20~$\mu$m) with the numbers in parentheses denoting the nominal monochromator exit-slit widths. The thick full lines represent the results from fits of the sum of four Voigt line profiles to the measured data. The four data sets were fitted simultaneously using a common set  of resonance parameters. The resulting fit values are listed in  Table~\ref{tab:Si1fit}. The individual line profiles are visualized by the thin full lines in panel d). Additional fit parameters were  $\Delta E$ (the Gaussian FWHM, $\pm$0.02~eV fit uncertainty in all cases), a constant background (practically negligible) and an overall energy shift amounting to a) $-0.19\pm$0.01~eV, b) $-0.150\pm0.007$~eV, c) $-0.089\pm0.005$~eV, and d) 0~eV.The shaded curves in panel d) is the present theoretical cross section $\sigma_{1,4}$ shifted in energy by $-1.7$~eV and convolved with a Gaussian with a FWHM of 0.3 eV.}
\end{figure}

The most prominent resonance feature in each absorption cross-section  is the lowest-energy resonance group, which is associated with $1s\to 3p$ excitation (Figure~\ref{fig:abs}). Its relative strength increases with increasing primary charge state. This also holds for the resonances associated with $1s$ excitation to higher $np$ subshells. In order to provide accurate resonance data we have measured the most prominent resonance features at lower photon-energy spreads. This was achieved at the expense of photon flux  by narrowing the monochromator exit-slit.

\begin{deluxetable}{lcccc}[b]
\tablecaption{\label{tab:Si1fit} Results of the resonance fits to the  Si$^+$   high-resolution data  ($\sigma_{1,4}$) for the $1s^{-1}\,3s^2\,3p^2$ resonance group (Figure~\ref{fig:Si1fit}). The resonance energies $E_r$, Lorentzian line widths $\Gamma$, and resonance strength $S_{1,4}$ were obtained from fitting Voigt line profiles to the experimental data. The column labelled by $S_\Sigma$ contains the derived absorption line strengths (see text). Numbers in parentheses denote the uncertainties that were obtained from the fit.
}
\tablehead{
    \colhead{Designation} &
    \colhead{$E_\mathrm{r}$}&
    \colhead{$\Gamma$}&
    \colhead{$S_{1,4}$} &
    \colhead{$S_\Sigma$}\\
    \colhead{} &
    \colhead{(eV)}&
    \colhead{(eV)}&
    \colhead{(Mb~eV)\tablenotemark{}} &
    \colhead{(Mb~eV)}
    }
\startdata
   $1s^{-1}\,3s^2\,3p^2\;^{2\!}P$ &  1841.26(1) & 0.38(22)  &  0.91(07) & 1.42(11)\\
   $1s^{-1}\,3s^2\,3p^2\;^{2\!}D$ &  1841.82(1) & 0.39(09)  &  0.70(14) & 1.09(22)\\
   $1s^{-1}\,3s^2\,3p^2\;^{4\!}P$ &  1842.35(5) & 0.34(13)   &  0.15(07) & 0.23(11)\\
   $1s^{-1}\,3s^2\,3p^2\;^{2\!}S$ &  1843.43(2) & 0.94(04)   &  0.26(01) & 0.41(02)\\
\enddata
\tablenotetext{}{1~Mb = $10^{-18}$~cm$^2$}
\end{deluxetable}

\begin{figure}[t]
\includegraphics[width=\linewidth]{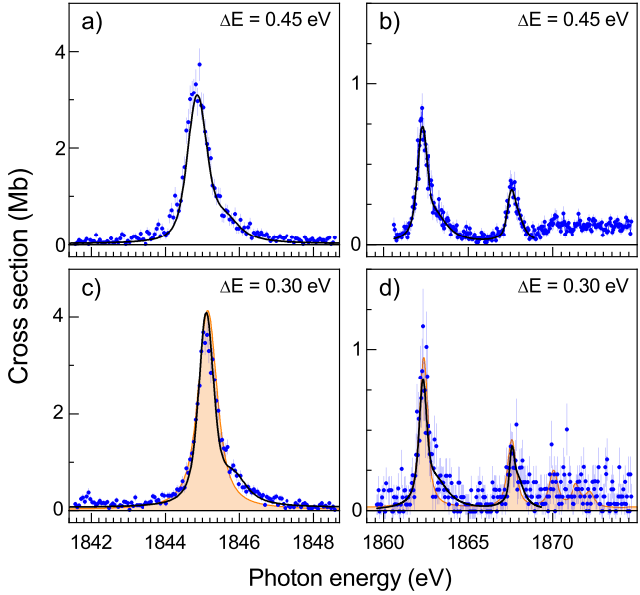}
\caption{\label{fig:Si2fit}Measured cross sections $\sigma_{2,4}$ for double photoionization of Si$^{2+}$ ions using different photon-energy spreads $\Delta E = 0.45$~eV  and $\Delta E =  0.30$~eV.  The thick full lines represent the results from fits of the sum of two (for a and c) and four (for b and d) Voigt line profiles to the measured data. The data sets in panels a) and c) were fitted simultaneously using a common set  of resonance parameters. The same holds for the data sets from panels b) and d). The resulting fit values are listed in  Table~\ref{tab:Si2fit}. In panel c), the individual line profiles are visualized by the thin full lines. Additional fit parameters were  $\Delta E$ (kept fixed), a constant background (subtracted) and an overall energy shift amounting to a) $-0.26\pm$0.02~eV, b) $-0.05\pm0.02$~eV, c) $0$~eV, and d) $0$~eV. The shaded curve in panels c) and d) is the present theoretical cross section $\sigma_{2,4}$ shifted in energy by $-1.8$~eV and convolved with a Gaussian with a FWHM of 0.3 eV.\\~\\}
\end{figure}

\begin{deluxetable}{lrrrr}[b]
\tablecaption{\label{tab:Si2fit}Results of the resonance fits to the  Si$^{2+}$   high-resolution data ($\sigma_{2,4}$) for the $1s^{-1}\,3s^2\,np$ resonance groups with $n=3,4,5$ (Figure~\ref{fig:Si2fit}). For the explanation of the notation see Table~\ref{tab:Si1fit}.
}
\tablehead{
    \colhead{Designation} &
    \colhead{$E_\mathrm{r}$}&
    \colhead{$\Gamma$}&
    \colhead{$S_{2,4}$} &
    \colhead{$S_\Sigma$}\\
    \colhead{} &
    \colhead{(eV)}&
    \colhead{(eV)}&
    \colhead{(Mb eV)} &
    \colhead{(Mb eV)}
    }
\startdata
 $1s^{-1}\,3s^2\,3p\;^{1\!}P$ &  1845.10(01) & 0.28(03)   &  2.48(16) & 3.33(22)  \\
 $1s^{-1}\,3s^2\,3p\;^{3\!}P$ &  1845.84(15) & 0.76(14)   &  0.72(16)  &  0.97(22)\\
  $1s^{-1}\,3s^2\,4p\;^{1\!}P$  &  1862.33(24) & 0.37(08)   &  0.55(10) & 0.60(11) \\
  $1s^{-1}\,3s^2\,4p\;^{3\!}P$  &  1863.10(20) & 1.77(37)   &  0.48(18) &  0.52(20) \\
  $1s^{-1}\,3s^2\,5p\;^{1\!}P$ &  1867.58(08) & 0.00(20)   &  0.07(05) &  \\
  $1s^{-1}\,3s^2\,5p\;^{3\!}P$ &  1867.85(08) & 0.84(15)   &  0.32(08) &  \\
\enddata
\end{deluxetable}

Figure~\ref{fig:Si1fit} displays high-resolution measurements of the Si$^+$($1s^{-1}\,3s^2\,3p^2$) resonance group, which consists of the four LS terms  $^2S$, $^2P$, $^4P$, and $^2D$ resulting from the coupling of the single $1s$ electron to the three $3p^2$ parent  terms $^1S$, $^3P$, and $^1D$.  For these measurements the dominant triple-ionization channel ($\sigma_{1,4}$) was chosen. The experimental resolving power increases when going from panel a) to panel d).   The numerical values of the respective photon energy spreads $\Delta E$ were obtained by a simultaneous fit of four Voigt line profiles to the four measured data sets. The resonance parameters that resulted from the fit,  i.e., the resonance energies $E_r$, the Lorentzian  line widths $\Gamma$, and the resonance strengths $S_{1,4}$ are listed in Table~\ref{tab:Si1fit}.

\begin{figure}[t]
\includegraphics[width=\linewidth]{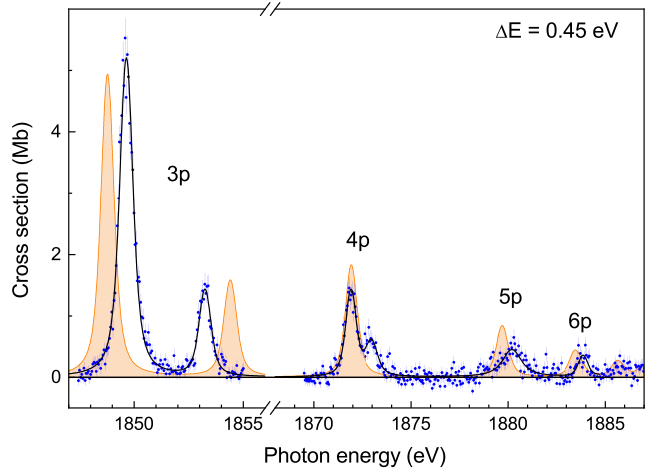}
\caption{\label{fig:Si3fit}Measured (symbols) and presently calculated (shaded curve) cross section $\sigma_{3,5}$ for double photoionization of Si$^{3+}$ ions using a  photon-energy spread $\Delta E = 0.45$~eV. The resonances are associated with the excitation of an $1s$ electron to the $3p$, $4p$, $5p$, and $6p$ subshells. The thick full line represents the result from a fit of the sum of six Voigt line profiles to the measured data.  The resulting fit values are listed in  Table~\ref{tab:Si3fit}. Additional fit parameters were  $\Delta E$ (kept fixed) and a constant background (subtracted from the data). The shaded curve is the present theoretical cross section $\sigma_{3,5}$ shifted in energy by $-1.2$~eV and convoluted with a Gaussian with a FWHM of 0.45 eV.}
\end{figure}

\begin{deluxetable}{lrrrr}
\tablecaption{\label{tab:Si3fit} Results of the resonance fits to the  Si$^{3+}$   high-resolution data ($\sigma_{3,5}$) for the $1s^{-1}\,3s\,np$ resonance groups with $n=3,4,5$ (Figure~\ref{fig:Si3fit}). For the explanation of the notation see Table~\ref{tab:Si1fit}.
}
\tablehead{
    \colhead{Designation} &
    \colhead{$E_\mathrm{r}$}&
    \colhead{$\Gamma$}&
    \colhead{$S_{3,5}$} &
    \colhead{$S_\Sigma$}\\
    \colhead{} &
    \colhead{(eV)}&
    \colhead{(eV)}&
    \colhead{(Mb eV)} &
    \colhead{(Mb eV)}
    }
\startdata
 $1s^{-1}\,3s\,3p\,(^{3\!}P)\;^{2\!}P$ &  1849.64(01) & 0.45(02)   &  5.20(09) &  5.20(09)\\
 $1s^{-1}\,3s\,3p\,(^{1\!}P)\;^{2\!}P$ &  1853.24(01) & 0.35(04)   &  1.23(06) &  1.23(06)\\
 $1s^{-1}\,3s\,4p\,(^{3\!}P)\;^{2\!}P$ &  1871.91(02) & 0.34(05)   &  1.19(08) &  1.19(08)\\
 $1s^{-1}\,3s\,4p\,(^{1\!}P)\;^{2\!}P$ &  1872.96(04) & 0.45(12)   &  0.56(08) &  0.56(08)\\
 $1s^{-1}\,3s\,5p\;^{2\!}P$ &  1880.20(05) & 1.00(44)   &  0.79(06) &  0.79(06)\\
 $1s^{-1}\,3s\,6p\;^{2\!}P$ &  1883.86(04) & 0.23(12)   &  0.25(04) &  0.25(04)\\
\enddata
\end{deluxetable}

In the fit, an overall energy shift was used as an additional fit parameter individually for each data set (see caption of Figure~\ref{fig:Si1fit}). The maximum shift amounts to $-0.19$~eV which is much less than the 1~eV uncertainty of the experimental photon-energy scale. Ideally, one would expect that the resonance positions were independent of the width of the monochromator's exit-slit. We attribute the observed shifts to a slight asymmetry of the photon-energy distribution and to mechanical imperfections of the photon-beamline's mechanics. A pertaining 0.2~eV uncertainty has already been considered in the error budget of our photon-energy calibration (see Section~\ref{sec:cal}).

In analogy to the work of  \citet{Schlachter2004a} for carbon, we note that the $1s^{-1}\,3s^2\,3p^2\;^{2\!}P$ and $1s^{-1}\,3s^2\,3p^2\;^{4\!}P$ levels, which are populated by $1s\to3p$ excitation of Si$^+$, can also be prepared by the removal of a $1s$ electron from  neutral  ground-term  neutral Si($1s^2\,3s^2\,3p^2\;^{3\!}P$). Therefore, the widths of the $1s^{-1}\,3s^2\,3p^2\;^{2\!}P$  and $1s^{-1}\,3s^2\,3p^2\;^{4\!}P$  levels from Table~\ref{tab:Si1fit} correspond to  the core-hole lifetime $\tau$ of initially neutral silicon. Both widths agree within their experimental uncertainties. From the less uncertain value we obtain $\tau=h/\Gamma = 1.4\pm0.5$~fs in agreement with the present calculations and with the value of 1.47~fs calculated by \citet{Palmeri2008}.

 In  Figure~\ref{fig:Si1fit}d, the present theoretical results for $\sigma_{1,4}$ are compared with the experimental high-resolution results. For the comparison, the theoretical resonances were shifted in energy by $-1.7$~eV and convoluted with a Gaussian with a FWHM of 0.3~eV. This resolution is a factor of 10 higher as compared to the one in Figure~\ref{fig:abs} and reveals some discrepancies between experiment and theory which are missed when looked at with larger photon-energy spread. The calculated splittings between the various $1s^{-1}\,3s^2\,3p^2$ terms are larger than what is found experimentally.

 The computed resonance strengths of individual terms contributing to $\sigma_{1,4}$ (see Figure \ref{fig:Si1fit}d) are larger than the experimental ones by about 50\%. Since the present theoretical  and experimental absorption cross-sections $\sigma_{1,\Sigma}$ agree with one another within the experimental uncertainty (Figure~\ref{fig:abs}b) it must be concluded that the mismatch for $\sigma_{1,4}$  is due to the overestimation of the corresponding branching ratios for the $1s^{-1}\,3s^2\,3p^2$ resonances.

 The theoretical branching ratios for the triple-ionization channel are almost the same for all of the $1s^{-1}\,3s^2\,3p^2$ resonances, i.e., their differences are smaller than the uncertainties of the individual $S_{1,4}$ values. This allows us to scale the fitted individual resonance strengths to the absorption cross-section. From a fit to the  $1s^{-1}\,3s^2\,3p^2$ peak in Figure~\ref{fig:abs}b we obtain a strength of 3.15(17)~Mb~eV for the sum of the four tabulated resonances. The sum of the $S_{1,4}$ values in Table~\ref{tab:Si1fit} amounts to 2.02(17)~Mb~eV.  Using these values we have calculated the tabulated  absorption resonance-strengths as $S_\Sigma = S_{1,4}\times3.15/2.02$. In the same way we have calculated $S_\Sigma$ separately for the $1s^{-1}\,3s^2\,3p$ [factor 4.30(3)/3.20(22)] and $1s^{-1}\,3s^2\,4p$ [factor (1.12(4)/1.03(21)] resonances of the Si$^{2+}$ absorption spectrum (Table~\ref{tab:Si2fit}). For Si$^{3+}$, the absorption spectrum is essentially identical with $\sigma_{3,5}$ and, thus, $S_\Sigma = \sigma_{3,5}$ for all resonances in Table~\ref{tab:Si3fit}.

 The fits to the Si$^{2+}$ and Si$^{3+}$ high-resolution data are shown in  Figures~\ref{fig:Si2fit} and \ref{fig:Si3fit}, respectively.  The agreement between experiment and the present theory (shifted by $-1.8$~eV for Si$^{2+}$ and by $-1.2$~eV for Si$^{3+}$, see Figure~\ref{fig:abs}) is remarkable, particularly, for  $\sigma_{2,4}$ (Figure \ref{fig:Si2fit}). There is less agreement for Si$^{3+}$, as already noted in the low-resolution data in Figures \ref{fig:all} and \ref{fig:abs}. The too large separation of the first two peaks is probably caused by subtleties concerning  the atomic fine structure associated with the angular momentum coupling of three open atomic subshells.

\subsection{Comparison with other forms of silicon}

\begin{figure}[t]
\includegraphics[width=\linewidth]{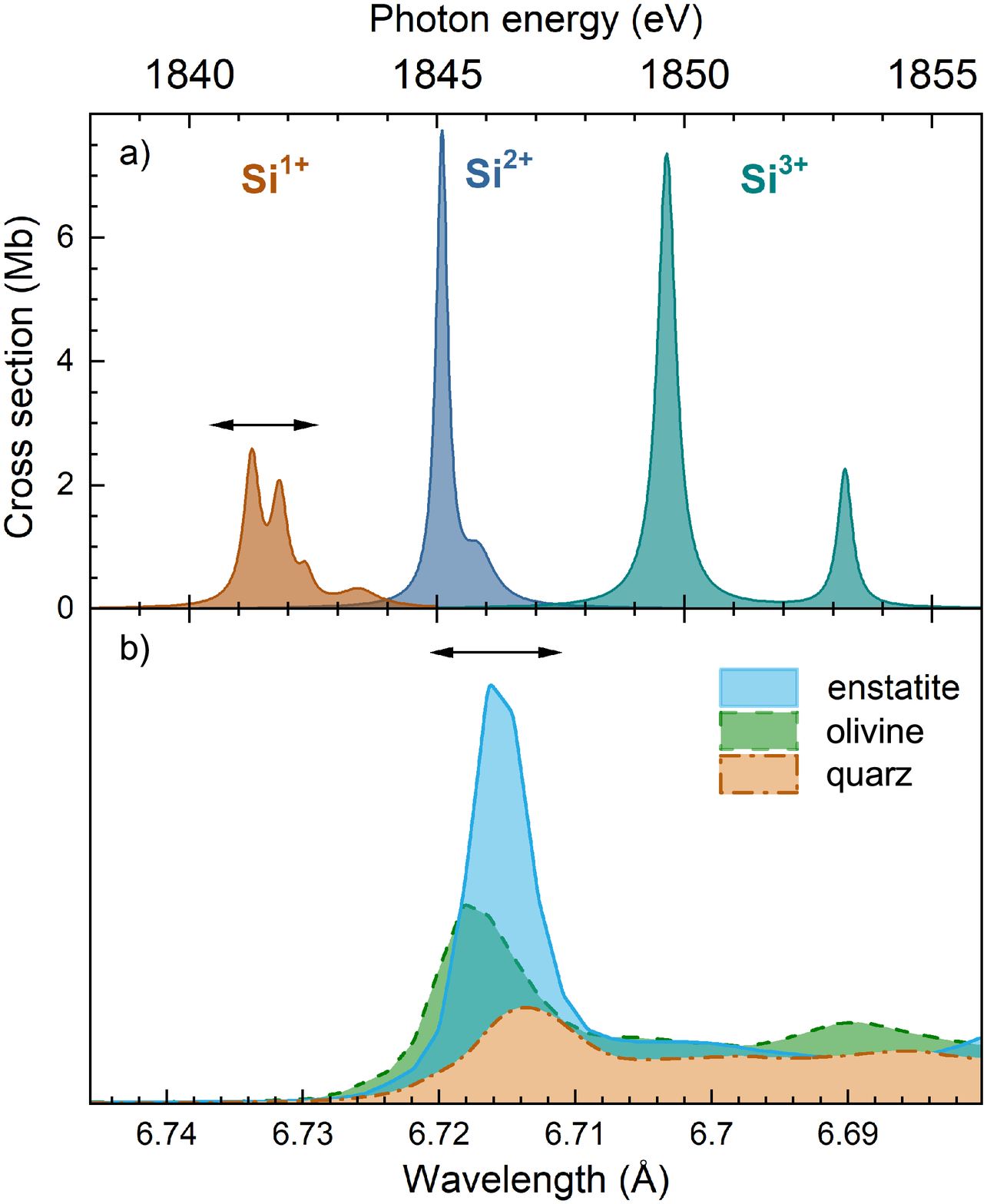}
\caption{\label{fig:comp} a)  Present results for the Si$^+$, Si$^{2+}$, and Si$^{3+}$ absorption features that are associated with $1s\to3p$ excitations. Each individual resonance is represented as a Lorentzian. The corresponding resonance parameters $E_r$, $\Gamma$ and $S_\Sigma$ are provided in Tables \ref{tab:Si1fit}, \ref{tab:Si2fit}, and \ref{tab:Si3fit}.   b) Absorption cross-sections of some silicon containing crystalline minerals \citep{Zeegers2019}. The horizontal arrows denote the experimental uncertainties of the photon energy scales amounting to a) $\pm1$~eV or $\pm0.0036$~\AA\  and b) $\pm1.5$~eV or $\pm0.0055$~\AA\ (see text).}
\end{figure}

Figure \ref{fig:comp} shows a comparison between the strongest Si $K$-shell absorption features for Si$^{+}$, Si$^{2+}$, and Si$^{3+}$ (panel a) with the absorption by  some silicon containing minerals that potentially occur in interstellar dust \citep[panel b,][]{Zeegers2019}.  In order to be able to discriminate between the different forms of silicon in x-ray absorption spectra, the energies of the absorption features must be known with sufficient precision.  The systematic uncertainty of the present experimental energy scale is $\pm1$~eV (Section~\ref{sec:cal}).  \citet{Zeegers2019}  did not quantify the uncertainty of their energy scale. Instead, they referred for their energy calibration to previous work \citep{Li1995,Nakanishi2009} which, in turn, is based on earlier investigations. The energy scale of \citet{Nakanishi2009}  can be traced back to the work of \citet{Wong1999} who mention that their energy calibration varies by 2--3~eV depending on the operating conditions of their synchrotron light source. This suggests that the uncertainty of the energy scale of \citet[][Figure~\ref{fig:comp}b]{Zeegers2019}  is at least $\pm 1.5$~eV.

Different minerals exhibit different chemical shifts of the Si $K$-shell absorption features which  vary by up to 1.5~eV \citep{Li1995}. This is still less than the combined energy uncertainties of the present energy scale and the one of \citet{Zeegers2019}. Therefore, we conclude that, in general, x-ray absorption spectroscopy can discriminate between absorption by gaseous silicon and absorption by silicon containing minerals. In particular, the Si$^+$  absorption feature in Figure~\ref{fig:comp}a appears at an unambiguous location ($\sim1841.5$~eV). This is also true for the corresponding absorption feature of Si$^-$ which occurs at an even lower energy of 1838.4~eV \citep[Figure~\ref{fig:abs}a,][]{Perry-Sassmannshausen2021}. To the best of our knowledge, there are no such experimental data for gaseous neutral silicon. In view of the findings for Si$^-$ and Si$^+$ its dominating absorption feature can be expected close to 1840~eV.

\citet{Gatuzz2020} investigated the gaseous component of the ISM using the theoretical absorption data of \citet{Witthoeft2011}. In their data, the $1s\to3p$ resonance group for neutral silicon is located at $\sim$1839.5~eV in accord with the above considerations. For the other charge states there are significant discrepancies  as already noted above. The deviations between  the theoretical resonance positions of \citet{Witthoeft2011} and the present experimental ones are 6.7, 4.4, and 9.1~eV for Si$^+$, Si$^{2+}$, and Si$^{3+}$, respectively (Figure~\ref{fig:abs}). All these differences are substantially larger than the uncertainty of the experimental energy scale and they are as large or larger than the energy differences between adjacent charge states (Figure~\ref{fig:comp}a). This will lead to a wrong assignment of astronomically observed resonance features if their analysis is based on these theoretical absorption cross sections.

\section{Conclusions and outlook}\label{sec:conc}

Using the photon-ion merged-beams technique at a synchrotron light source we have measured cross sections for multiple photoionization of low-charged Si$^+$, Si$^{2+}$, and Si$^{3+}$ ions and derived precise absorption data (resonance positions, widths, and strengths) which can be directly used for the astrophysical modelling of the silicon $K$-shell absorption by interstellar gas clouds and other cosmic objects as well as for the benchmarking the theoretical calculations. From the widths of the core excited resonances in Si$^+$ we inferred a value for the core-hole lifetime of $1s$-ionized neutral silicon of $1.4\pm0.5$~fs.

Previously theoretically predicted absorption features deviate significantly in energy from the present experimental findings, whereas the present large-scale MCDHF calculations agree much better with the experimental results. In addition, the present computations also capture the hole-deexcitation cascades following the initial creation of a $K$-shell hole. The obtained product charge-state distributions, which are required for an accurate modelling of the charge balance in astrophysical plasmas,  agree remarkably well with the experimental results, despite of the simplifications that had to be applied to keep the calculations tractable. Current code development aims at systematically expanding such cascade calculations to improve the treatment of the deexcitation processes \citep{Fritzsche2019,Fritzsche2021}.

When comparing absorption data from different sources the calibration of the photon energy scales is an issue of concern.  The present experimental uncertainty of $\pm 1$ eV is sufficient for discriminating between absorption by gaseous and solid silicon-containing matter in the x-ray absorption spectra from the currently operated x-ray telescopes. The accuracy demands will increase with increasing resolving power of future missions such as Athena \citep{Barret2020}. Meeting these demands will require the development of more accurate calibration standards at synchrotron light sources. Promising candidates are  few-electron atomic ions which promise calibration uncertainties of less than 10~meV  \citep[see, e.g.,][]{Mueller2018c,Stierhof2022}. Corresponding activities are under way at the PIPE setup.

\begin{acknowledgments}
We acknowledge DESY (Hamburg, Germany), a member of the Helmholtz Association HGF, for the provision of experimental facilities. Parts of this research were carried out at PETRA\,III and we would like to thank Jens Buck, Moritz Hoesch, Frank Scholz, and J\"orn Seltmann for assistance in using beamline P04.
This work was supported by the German Federal Ministry for Education and Research (BMBF, grant numbers 05K16RG1, 05K16GUC, and 05K16SJA) and by the Deutsche Forschungsgemeinschaft (DFG, grant number 389115454).
\end{acknowledgments}


\begin{thebibliography}{}
\expandafter\ifx\csname natexlab\endcsname\relax\def\natexlab#1{#1}\fi
\providecommand{\url}[1]{\href{#1}{#1}}
\providecommand{\dodoi}[1]{doi:~\href{http://doi.org/#1}{\nolinkurl{#1}}}
\providecommand{\doeprint}[1]{\href{http://ascl.net/#1}{\nolinkurl{http://ascl.net/#1}}}
\providecommand{\doarXiv}[1]{\href{https://arxiv.org/abs/#1}{\nolinkurl{https://arxiv.org/abs/#1}}}

\bibitem[{Barret {et~al.}(2020)Barret, Decourchelle, Fabian, Guainazzi, Nandra,
  Smith, \& den Herder}]{Barret2020}
Barret, D., Decourchelle, A., Fabian, A., {et~al.} 2020, Astron. N., 341, 224,
  \dodoi{10.1002/asna.202023782}

\bibitem[{Beerwerth \& Fritzsche(2017)}]{Beerwerth2017}
Beerwerth, R., \& Fritzsche, S. 2017, EPJD, 71, 253,
  \dodoi{10.1140/epjd/e2017-80064-3}

\bibitem[{Beerwerth {et~al.}(2019)Beerwerth, Buhr, Perry-Sassmannshausen,
  Stock, Bari, Holste, Kilcoyne, Reinwardt, Ricz, Savin, Schubert, Martins,
  M{\"{u}}ller, Fritzsche, \& Schippers}]{Beerwerth2019}
Beerwerth, R., Buhr, T., Perry-Sassmannshausen, A., {et~al.} 2019, ApJ, 887,
  189, \dodoi{10.3847/1538-4357/ab5118}

\bibitem[{Bizau {et~al.}(2009)Bizau, Mosnier, Kennedy, Cubaynes, Wuilleumier,
  Blancard, Champeaux, \& Folkmann}]{Bizau2009}
Bizau, J.-M., Mosnier, J.-P., Kennedy, E.~T., {et~al.} 2009, PhRvA, 79, 033407

\bibitem[{Buth {et~al.}(2018)Buth, Beerwerth, Obaid, Berrah, Cederbaum, \&
  Fritzsche}]{Buth2018}
Buth, C., Beerwerth, R., Obaid, R., {et~al.} 2018, JPhB, 51, 055602,
  \dodoi{10.1088/1361-6455/aaa39a}

\bibitem[{Deslattes {et~al.}(2003)Deslattes, Kessler, Indelicato, de~Billy,
  Lindroth, \& Anton}]{Deslattes2003}
Deslattes, R.~D., Kessler, E.~G., Indelicato, J.~P., {et~al.} 2003, RvMP, 75,
  35, \dodoi{10.1103/RevModPhys.75.35}

\bibitem[{Fritzsche(2012)}]{Fritzsche2012a}
Fritzsche, S. 2012, CoPhC, 183, 1525, \dodoi{10.1016/j.cpc.2012.02.016}

\bibitem[{Fritzsche(2019)}]{Fritzsche2019}
---. 2019, CoPhC, 240, 1, \dodoi{10.1016/j.cpc.2019.01.012}

\bibitem[{Fritzsche {et~al.}(2021)Fritzsche, Palmeri, \&
  Schippers}]{Fritzsche2021}
Fritzsche, S., Palmeri, P., \& Schippers, S. 2021, Symmetry, 13, 520,
  \dodoi{10.3390/sym13030520}

\bibitem[{Froese~Fischer {et~al.}(2006)Froese~Fischer, Tachiev, \&
  Irimia}]{FroeseFischer2006}
Froese~Fischer, C., Tachiev, G., \& Irimia, A. 2006, ADNDT, 92, 607,
  \dodoi{10.1016/j.adt.2006.03.001}

\bibitem[{Gatuzz {et~al.}(2020)Gatuzz, Gorczyca, Hasoglu, Schulz, Corrales, \&
  Mendoza}]{Gatuzz2020}
Gatuzz, E., Gorczyca, T.~W., Hasoglu, M.~F., {et~al.} 2020, MNRAS, 498, L20,
  \dodoi{10.1093/mnrasl/slaa119}

\bibitem[{Hasoglu {et~al.}(2021)Hasoglu, Gorczyca, \& Manson}]{Hasoglu2021}
Hasoglu, M.~F., Gorczyca, T.~W., \& Manson, S.~T. 2021, PhyS, 96, 124024,
  \dodoi{10.1088/1402-4896/ac0b84}

\bibitem[{Henke {et~al.}(1993)Henke, Gullikson, \& Davis}]{Henke1993}
Henke, B.~L., Gullikson, E.~M., \& Davis, J.~C. 1993, ADNDT, 54, 181,
  \dodoi{10.1006/adnd.1993.1013}

\bibitem[{Howald {et~al.}(1986)Howald, Gregory, Meyer, Phaneuf, M\"{u}ller,
  Djuri\'{c}, \& Dunn}]{Howald1986a}
Howald, A.~M., Gregory, D.~C., Meyer, F.~W., {et~al.} 1986, PhRvA, 33, 3779,
  \dodoi{10.1103/PhysRevA.33.3779}

\bibitem[{Ibuki {et~al.}(2002)Ibuki, Okada, Kamimori, Sasaki, Yoshida, Hiraya,
  Suzuki, Saito, Nagaoka, Shimizu, Ohashi, \& Tamenori}]{Ibuki2002}
Ibuki, T., Okada, K., Kamimori, K., {et~al.} 2002, SRL,
  09, 85, \dodoi{10.1142/S0218625X02001987}

\bibitem[{Jenkins(2009)}]{Jenkins2009}
Jenkins, E.~B. 2009, ApJ, 700, 1299, \dodoi{10.1088/0004-637X/700/2/1299}

\bibitem[{J\"{o}nsson {et~al.}(2007)J\"{o}nsson, He, Froese-Fischer, \&
  Grant}]{Joensson2007a}
J\"{o}nsson, P., He, X., Froese-Fischer, C., \& Grant, I.~P. 2007, CoPhC, 177,
  597, \dodoi{10.1016/j.cpc.2007.06.002}

\bibitem[{Kato {et~al.}(2007)Kato, Morishita, Oura, Yamaoka, Tamenori, Okada,
  Matsudo, Gejo, Suzuki, \& Saito}]{Kato2007}
Kato, M., Morishita, Y., Oura, M., {et~al.} 2007, AIP Conference Proceedings,
  879, 1121, \dodoi{10.1063/1.2436260}

\bibitem[{Kennedy {et~al.}(2014)Kennedy, Mosnier, Van~Kampen, Cubaynes,
  Guilbaud, Blancard, McLaughlin, \& Bizau}]{Kennedy2014}
Kennedy, E.~T., Mosnier, J.-P., Van~Kampen, P., {et~al.} 2014, PhRvA, 90,
  063409, \dodoi{10.1103/PhysRevA.90.063409}

\bibitem[{Ku\v{c}as {et~al.}(2012)Ku\v{c}as, Karazija, \&
  Momkauskait\.{e}}]{Kucas2012}
Ku\v{c}as, S., Karazija, R., \& Momkauskait\.{e}, A. 2012, ApJ, 750, 90,
  \dodoi{10.1088/0004-637X/750/2/90}

\bibitem[{Ku\v{c}as {et~al.}(2015)Ku\v{c}as, Momkauskait\.{e}, \&
  Karazija}]{Kucas2015}
Ku\v{c}as, S., Momkauskait\.{e}, A., \& Karazija, R. 2015, ApJ, 810, 26,
  \dodoi{10.1088/0004-637X/810/1/26}

\bibitem[{Li {et~al.}(1995)Li, Bancroft, Fleet, \& Feng}]{Li1995}
Li, D., Bancroft, G.~M., Fleet, M.~E., \& Feng, X.~H. 1995, PCM,
  22, 115, \dodoi{10.1007/BF00202471}

\bibitem[{Mosnier {et~al.}(2003)Mosnier, Sayyad, Kennedy, Bizau, Cubaynes,
  Wuilleumier, Champeaux, Blancard, Varma, Banerjee, Deshmukh, \&
  Manson}]{Mosnier2003a}
Mosnier, J.-P., Sayyad, M.~H., Kennedy, E.~T., {et~al.} 2003, PhRvA, 68,
  052712, \dodoi{10.1103/PhysRevA.68.052712}

\bibitem[{M\"{u}ller {et~al.}(2015)M\"{u}ller, {Borovik~Jr.}, Buhr, Hellhund,
  Holste, Kilcoyne, Klumpp, Martins, Ricz, Viefhaus, \&
  Schippers}]{Mueller2015a}
M\"{u}ller, A., {Borovik~Jr.}, A., Buhr, T., {et~al.} 2015, PhRvL, 114, 013002,
  \dodoi{10.1103/PhysRevLett.114.013002}

\bibitem[{M\"{u}ller {et~al.}(2017)M\"{u}ller, Bernhardt, {Borovik~Jr.}, Buhr,
  Hellhund, Holste, Kilcoyne, Klumpp, Martins, Ricz, Seltmann, Viefhaus, \&
  Schippers}]{Mueller2017}
M\"{u}ller, A., Bernhardt, D., {Borovik~Jr.}, A., {et~al.} 2017, ApJ, 836, 166,
  \dodoi{10.3847/1538-4357/836/2/166}

\bibitem[{M\"{u}ller {et~al.}(2018)M\"{u}ller, Lindroth, Bari, {Borovik~Jr.},
  Hillenbrand, Holste, Indelicato, Kilcoyne, Klumpp, Martins, Viefhaus,
  Wilhelm, \& Schippers}]{Mueller2018c}
M\"{u}ller, A., Lindroth, E., Bari, S., {et~al.} 2018, PhRvA, 98, 033416,
  \dodoi{10.1103/PhysRevA.98.033416}

\bibitem[{M\"uller {et~al.}(2021)M\"uller, Martins, Borovik, Buhr,
  Perry-Sassmannshausen, Reinwardt, Trinter, Schippers, Fritzsche, \&
  Kheifets}]{Mueller2021b}
M\"uller, A., Martins, M., Borovik, A., {et~al.} 2021, PhRvA, 104, 033105,
  \dodoi{10.1103/PhysRevA.104.033105}

\bibitem[{Nagaoka {et~al.}(2000)Nagaoka, Ibuki, Saito, Shimizu, Senba,
  Kamimori, Tamenori, Ohashi, \& Suzuki}]{Nagaoka2000}
Nagaoka, S., Ibuki, T., Saito, N., {et~al.} 2000, JPhB, 33, L605,
  \dodoi{10.1088/0953-4075/33/17/102}

\bibitem[{Nakanishi \& Ohta(2009)}]{Nakanishi2009}
Nakanishi, K., \& Ohta, T. 2009, JPCM., 21, 104214,
  \dodoi{10.1088/0953-8984/21/10/104214}

\bibitem[{Okada {et~al.}(2005)Okada, Kosugi, Fujii, Nagaoka, Ibuki, Samori,
  Tamenori, Ohashi, Suzuki, \& Ohno}]{Okada2005}
Okada, K., Kosugi, M., Fujii, A., {et~al.} 2005, JPhB, 38, 421,
  \dodoi{10.1088/0953-4075/38/4/009}

\bibitem[{Palmeri {et~al.}(2008)Palmeri, Quinet, Mendoza, Bautista, Garc\'{i}a,
  \& Kallman}]{Palmeri2008}
Palmeri, P., Quinet, P., Mendoza, C., {et~al.} 2008, ApJS, 177, 408,
  \dodoi{10.1086/587804}

\bibitem[{Perry-Sassmannshausen {et~al.}(2021)Perry-Sassmannshausen, Buhr,
  Martins, Reinwardt, Trinter, M{\"{u}}ller, Fritzsche, \&
  Schippers}]{Perry-Sassmannshausen2021}
Perry-Sassmannshausen, A., Buhr, T., Martins, M., {et~al.} 2021, PhRvA, 104,
  053107, \dodoi{10.1103/PhysRevA.104.053107}

\bibitem[{Perry-Sassmannshausen {et~al.}(2020)Perry-Sassmannshausen, Buhr,
  {Borovik~Jr.}, Martins, Reinwardt, Ricz, Stock, Trinter, M\"uller, Fritzsche,
  \& Schippers}]{Perry-Sassmannshausen2020}
Perry-Sassmannshausen, A., Buhr, T., {Borovik~Jr.}, A., {et~al.} 2020, PhRvL,
  124, 083203, \dodoi{10.1103/PhysRevLett.124.083203}

\bibitem[{Rogantini {et~al.}(2020)Rogantini, Costantini, Zeegers, Mehdipour,
  Psaradaki, Raassen, de~Vries, \& Waters}]{Rogantini2020}
Rogantini, D., Costantini, E., Zeegers, S.~T., {et~al.} 2020, 641, A149,
  \dodoi{10.1051/0004-6361/201936805}

\bibitem[{Schippers {et~al.}(2016{\natexlab{a}})Schippers, Kilcoyne, Phaneuf,
  \& M\"{u}ller}]{Schippers2016}
Schippers, S., Kilcoyne, A. L.~D., Phaneuf, R.~A., \& M\"{u}ller, A.
  2016{\natexlab{a}}, ConPh, 57, 215, \dodoi{10.1080/00107514.2015.1109771}

\bibitem[{Schippers \& M\"uller(2020)}]{Schippers2020c}
Schippers, S., \& M\"uller, A. 2020, Atoms, 8, 45, \dodoi{10.3390/atoms8030045}

\bibitem[{Schippers {et~al.}(2014)Schippers, Ricz, Buhr, {Borovik~Jr.},
  Hellhund, Holste, Huber, Sch\"{a}fer, Schury, Klumpp, Mertens, Martins,
  Flesch, Ulrich, R\"{u}hl, Jahnke, Lower, Metz, Schmidt, Sch\"{o}ffler,
  Williams, Glaser, Scholz, Seltmann, Viefhaus, Dorn, Wolf, Ullrich, \&
  M\"{u}ller}]{Schippers2014}
Schippers, S., Ricz, S., Buhr, T., {et~al.} 2014, JPhB, 47, 115602,
  \dodoi{10.1088/0953-4075/47/11/115602}

\bibitem[{Schippers {et~al.}(2016{\natexlab{b}})Schippers, Beerwerth, Abrok,
  Bari, Buhr, Martins, Ricz, Viefhaus, Fritzsche, \&
  M\"{u}ller}]{Schippers2016a}
Schippers, S., Beerwerth, R., Abrok, L., {et~al.} 2016{\natexlab{b}}, PhRvA,
  94, 041401(R), \dodoi{10.1103/PhysRevA.94.041401}

\bibitem[{Schippers {et~al.}(2017)Schippers, Martins, Beerwerth, Bari, Holste,
  Schubert, Viefhaus, Savin, Fritzsche, \& M\"{u}ller}]{Schippers2017}
Schippers, S., Martins, M., Beerwerth, R., {et~al.} 2017, ApJ, 849, 5,
  \dodoi{10.3847/1538-4357/aa8fcc}

\bibitem[{Schippers {et~al.}(2020)Schippers, Buhr, {Borovik~Jr.}, Holste,
  Perry-Sassmannshausen, Mertens, Reinwardt, Martins, Klumpp, Schubert, Bari,
  Beerwerth, Fritzsche, Ricz, Hellhund, \& M\"{u}ller}]{Schippers2020}
Schippers, S., Buhr, T., {Borovik~Jr.}, A., {et~al.} 2020, XRS, 49,
  11, \dodoi{10.1002/xrs.3035}

\bibitem[{Schippers {et~al.}(2021)Schippers, Beerwerth, Bari, Buhr, Holste,
  Kilcoyne, Perry-Sassmannshausen, Phaneuf, Reinwardt, Savin, Schubert,
  Fritzsche, Martins, \& M{\"{u}}ller}]{Schippers2021}
Schippers, S., Beerwerth, R., Bari, S., {et~al.} 2021, ApJ, 908, 52,
  \dodoi{10.3847/1538-4357/abcc64}

\bibitem[{Schlachter {et~al.}(2004)Schlachter, Sant{'}Anna, Covington, Aguilar,
  Gharaibeh, Emmons, Scully, Phaneuf, Hinojosa, \'{A}lvarez, Cisneros,
  M\"{u}ller, \& McLaughlin}]{Schlachter2004a}
Schlachter, A.~S., Sant{'}Anna, M.~M., Covington, A.~M., {et~al.} 2004, JPhB,
  37, L103, \dodoi{10.1088/0953-4075/37/5/L03}

\bibitem[{Schmelz {et~al.}(1995)Schmelz, Gaveau, Reynaud, Heinzel,
  Baumg{\"{a}}rtel, \& R{\"{u}}hl}]{Schmelz1995}
Schmelz, H.~C., Gaveau, M.~A., Reynaud, C., {et~al.} 1995, PhyB, 208-209,
  519, \dodoi{https://doi.org/10.1016/0921-4526(94)01038-3}

\bibitem[{Schmidt {et~al.}(2007)Schmidt, Bernhardt, M\"{u}ller, Schippers,
  Fritzsche, Hoffmann, Jaroshevich, Krantz, Lestinsky, Orlov, Wolf, Luki\'{c},
  \& Savin}]{Schmidt2007b}
Schmidt, E.~W., Bernhardt, D., M\"{u}ller, A., {et~al.} 2007, PhRvA, 76,
  032717, \dodoi{10.1103/PhysRevA.76.032717}

\bibitem[{Stierhof {et~al.}(2022)Stierhof, K{\"{u}}hn, Winter, Micke,
  Steinbr{\"{u}}gge, Shah, Hell, Bissinger, Hirsch, Ballhausen, Lang,
  Gr{\"{a}}fe, Wipf, Cumbee, Betancourt-Martinez, Park, Niskanen, Chung,
  Porter, St{\"{o}}hlker, Pfeifer, Brown, Bernitt, Hansmann, Wilms, Crepso
  L{\'{o}}pez-Urrutia, \& Leutenegger}]{Stierhof2022}
Stierhof, J., K{\"{u}}hn, S., Winter, M., {et~al.} 2022, EPJD, 76, 38,
  \dodoi{10.1140/epjd/s10053-022-00355-0}

\bibitem[{Stock {et~al.}(2017)Stock, Beerwerth, \& Fritzsche}]{Stock2017}
Stock, S., Beerwerth, R., \& Fritzsche, S. 2017, PhRvA, 95, 053407,
  \dodoi{10.1103/PhysRevA.95.053407}

\bibitem[{Tiesinga {et~al.}(2021)Tiesinga, Mohr, Newell, \&
  Taylor}]{Tiesinga2021}
Tiesinga, E., Mohr, P.~J., Newell, D.~B., \& Taylor, B.~N. 2021, RvMP, 93,
  025010, \dodoi{10.1103/RevModPhys.93.025010}

\bibitem[{Verner {et~al.}(1993)Verner, Yakovlev, Band, \&
  Trzhaskovskaya.}]{Verner1993a}
Verner, D.~A., Yakovlev, D.~G., Band, I.~M., \& Trzhaskovskaya., M.~B. 1993,
  ADNDT, 55, 233, \dodoi{10.1006/adnd.1993.1022}

\bibitem[{Viefhaus {et~al.}(2013)Viefhaus, Scholz, Deinert, Glaser, Ilchen,
  Seltmann, Walter, \& Siewert}]{Viefhaus2013}
Viefhaus, J., Scholz, F., Deinert, S., {et~al.} 2013, NIMPA, 710, 151,
  \dodoi{10.1016/j.nima.2012.10.110}

\bibitem[{Witthoeft {et~al.}(2011)Witthoeft, Garc\'{i}a, Kallman, Bautista,
  Mendoza, Palmeri, \& Quinet}]{Witthoeft2011}
Witthoeft, M.~C., Garc\'{i}a, J., Kallman, T.~R., {et~al.} 2011, ApJS, 1992, 7,
  \dodoi{10.1088/0067-0049/192/1/7}

\bibitem[{Wong {et~al.}(1999)Wong, Tanaka, Rowen, Sch{\"{a}}fers, M{\"{u}}ller,
  \& Rek}]{Wong1999}
Wong, J., Tanaka, T., Rowen, M., {et~al.} 1999, J Snychr. Rad., 6, 1086,
  \dodoi{10.1107/S0909049599009000}

\bibitem[{Wuilleumier(1971)}]{Wuilleumier1971}
Wuilleumier, F. 1971, J. Physique, 32, C4.88, \dodoi{10.1051/jphyscol:1971418}

\bibitem[{Zeegers {et~al.}(2019)Zeegers, Costantini, Rogantini, de~Vries,
  Mutschke, Mohr, de~Groot, \& Tielens}]{Zeegers2019}
Zeegers, S.~T., Costantini, E., Rogantini, D., {et~al.} 2019, 627, A16,
  \dodoi{10.1051/0004-6361/201935050}

\bibitem[{Zeegers {et~al.}(2017)Zeegers, Costantini, {de Vries}, Tielens,
  Chihara, de~Groot, Mutschke, Waters, \& Zeidler}]{Zeegers2017}
Zeegers, S.~T., Costantini, E., {de Vries}, C.~P., {et~al.} 2017, 599, A117,
  \dodoi{10.1051/0004-6361/201628507}

\end{thebibliography}

\end{document}